\documentclass{webofc}
\usepackage[varg]{txfonts}   % Web of Conferences font
\usepackage{amssymb}
\usepackage{graphicx}
\usepackage{paralist}
\usepackage{subfigure}
\usepackage[normalem]{ulem}
\usepackage{cancel}

\def\beq{\begin{equation}}
\def\eeq{\end{equation}}
\def\bea{\begin{eqnarray}}
\def\eea{\end{eqnarray}}

\newcommand{\GeV}{ {\rm GeV}}

\newcommand{\der}{\ensuremath{{\operatorname{d}}}}

\newcommand{\at}{\makeatletter @\makeatother}

\woctitle{POETIC VI: 6$^{th}$ International conference on Physics Opportunities at Electron-Ion collider}
\begin{document}
\title{Intrinsic bottom and its impact on heavy new physics at the LHC}

\author{Florian Lyonnet\inst{1}\fnsep\thanks{\email{flyonnet@smu.edu}}}

\institute{Southern Methodist University, Dallas, TX, USA
          }

\abstract{%
	Heavy quark parton distribution functions (PDFs) play an important role in several Standard Model and New Physics processes. Most analyses rely on the assumption that the charm and bottom PDFs are generated perturbatively by gluon splitting and do not involve any non-perturbative degrees of freedom. On the other hand, non- perturbative, intrinsic heavy quark parton distributions have been predicted in the literature. We demonstrate that to a very good approximation the scale-evolution of the intrinsic heavy quark content of the nucleon is governed by non-singlet evolution equations. This allows to analyze the intrinsic heavy quark distributions without having to resort to a full-fledged global analysis of parton distribution functions. We exploit this freedom to model intrinsic bottom distributions which are so far missing in the literature. We estimate the impact of the non-perturbative contribution to the charm and bottom-quark PDFs and on several important parton-parton luminosities at the LHC.
}
\maketitle

\section{Introduction}

Several Standard Model (SM) and New Physics (NP) processes at the CERN Large Hadron Collider (LHC) crucially depend on heavy quark parton distribution functions (PDFs); see for instance ~\cite{Maltoni:2012pa} for key processes involving the bottom quark PDF.

In the standard approach, the heavy quark distributions are generated {\em radiatively} via the DGLAP evolution equations, starting with a perturbatively calculable boundary condition. However, a purely perturbative, {\it extrinsic}, treatment for the heavy quarks might {\em not} be adequate; several models indeed postulate a non-perturbative, {\it intrinsic}, heavy quark component e.g. light-cone~\cite{Brodsky:1980pb,Brodsky:1981se} and meson cloud models~\cite{Navarra:1995rq,Paiva:1996dd,Melnitchouk:1997ig}.

Along the years, various global fits have been performed to estimate the amount of intrinsic charm (IC) allowed in the nucleon, see~\cite{Brodsky:2015fna} for a recent review.
Among them, the two most recent analyses~\cite{Gao:2013xoa,Jimenez-Delgado:2014zga} set significantly different limits on the allowed IC contribution highlighting the utility of the techniques discussed in this paper as we can freely adjust the amount of IC/IB contributions without having to regenerate a complete global analysis for each case. 

In this contribution, we summarize a technique introduced in~\cite{Lyonnet:2015dca} which can provide IB PDFs for any generic
non-IB PDF set. This approach relies on the fact that the intrinsic bottom PDF evolves (to an excellent precision) according to
a standalone non-singlet evolution equation. It is then easy to produce a matched set
of IB and non-IB PDFs. Note that obtaining information on the IB content of the nucleon from a global fit is doomed because the data entering global analyses of proton PDFs do not constrain the IB PDFs.

Sec.~\ref{sec:intrinsic}, demonstrates that to a good approximation the scale-evolution
of the intrinsic PDF is governed by a non-singlet evolution equation; it also provide a set of matched IC/IB PDFs.
The IB PDFs are then used in Sec.~\ref{sec:numerics} to obtain predictions for parton--parton luminosities relevant at the LHC.
Finally, our results and conclusions are summarized in Sec.~\ref{sec:conclusions}.

%%%%%%%%%%%%%%%%%%%%%%%%%%%%%%%%%%%%%%%%%%%%%
\section{Parton distribution functions for intrinsic heavy quarks} 
\label{sec:intrinsic}
%%%%%%%%%%%%%%%%%%%%%%%%%%%%%%%%%%%%%%%%%%%%%

%============================================
\subsection{Definition and Evolution}
\label{sec:definition}
%============================================
In the context of a global analysis of PDFs the different parton flavors are specified via a boundary condition at the input scale $\mu_0$ which is typically of the order $\mathcal{O}(1\ \mathrm{GeV})$. The PDFs at higher scales $\mu > \mu_0$ are then obtained by solving the DGLAP evolution equations with these boundary conditions. Thus, a non-perturbative (intrinsic) heavy quark distribution $Q_1$ can be defined at the input scale $\mu_0$ as the difference of the full boundary condition for the heavy quark PDF $Q$ and the perturbatively calculable (extrinsic) boundary condition $Q_0$.

The Dokshitzer-Gribov-Lipatov-Altarelli-Parisi (DGLAP) evolution equations read
$\dot f_i=P_{ij}\otimes f_j$, where $f_i=\begin{pmatrix} g\\q\\b \end{pmatrix}$
\begin{eqnarray}
\label{eq:DGLAP2a}
\dot g &=& P_{gg}\otimes g+P_{gq}\otimes q+P_{gQ}\otimes Q_0+ {\cancel{P_{gQ}\otimes Q_1}}\, ,
 \\
\label{eq:DGLAP2b}
\dot q &=& P_{qg}\otimes g+P_{qq}\otimes q+P_{qQ}\otimes Q_0+ {\cancel{P_{qQ}\otimes Q_1}}\, ,
 \\
\label{eq:DGLAP2c}
\dot Q_0 + \dot Q_1 &=& P_{Qg}\otimes g+P_{Qq}\otimes q+P_{QQ}\otimes Q_0+ P_{QQ}\otimes Q_1 \, ,
\end{eqnarray}
where vector of light quarks is denoted `$q$' and the heavy quark distribution by `$Q$' (where $Q=c$ or $Q=b$). Note that we substituted $Q=Q_0+Q_1$ where $Q_0$ denotes the usual radiatively generated extrinsic heavy quark component and $Q_1$ is the non-perturbative intrinsic heavy quark distribution.\footnote{Strictly speaking, the decomposition of $Q$ into $Q_0$ and $Q_1$ is defined at the input scale where the calculable boundary condition for $Q_0$ is known. Consequently, $Q_1:=Q-Q_0$ is known as well. Only due to the approximations in Eqs.\ \protect\eqref{eq:DGLAP2a} and \protect\eqref{eq:DGLAP2b} it is possible to entirely decouple $Q_0$ from $Q_1$ so that the decomposition becomes meaningful at any scale.}

If one neglects the tiny contribution coming from the crossed out terms the system of evolution equations can be separated into two independent parts.

For the system of gluon, light quarks and extrinsic heavy quark ($g,q,Q_{0}$) one recovers the same evolution equations as in the standard approach without an intrinsic heavy quark component while the intrinsic heavy quark distribution, $Q_1$, follows a standalone non-singlet evolution equation, $\dot Q_1 = P_{QQ}\otimes Q_1 \, .$ 

It is then apparent that it is possible to entirely decouple the analysis of the intrinsic heavy quark distribution from the rest of the system by allowing  for a small violation of the sum rule. Therefore, PDFs for the gluon, the light quarks and the extrinsic heavy quark can be taken from a global analysis in the standard approach where they already saturate the momentum sum rule. On top of these PDFs the intrinsic heavy quark PDF can be determined in a standalone analysis using the non-singlet evolution equation. 

The violation of the momentum sum rule induced by the term $\int_0^1\ \der x\  x\  \left(Q_1 + \bar{Q}_1\right)$ is, however, very small for bottom quarks\footnote{It is also acceptable in case of charm provided that the allowed normalization of IC is not too big.}. Numerical checks of the validity of our approximations are performed below. 
%
%============================================
\subsection{Boundary condition for intrinsic heavy quarks}
\label{sec:bc}
%============================================

The $x$-dependence of the intrinsic charm (IC) parton distribution function is predicted by the BHPS model~\cite{Brodsky:1980pb}to be:
\begin{equation}
c_1(x) = \bar c_1(x) \propto x^2 [6 x (1+x) \ln x + (1-x)(1+10 x+x^2)]\, .
\label{eq:bhps}
\end{equation}
We expect the $x$-shape of the intrinsic bottom distribution $b_1(x)$ to be very similar to the one of the intrinsic charm distribution. Furthermore, the normalization of IB is expected to be parametrically suppressed with respect to IC by a factor $m_c^2/m_b^2 \simeq 0.1$. To fix the freedom related to the scale of the boundary condition we use in the following the {\it Same Scales} boundary condition, which remains valid at any scale $Q$: $b_1(x,m_c) = \frac{m_c^2}{m_b^2} c_1(x,m_c)\, .$ Let us also note that in this approach it would be no problem to work with asymmetric boundary conditions, $\bar c_1(x) \ne c_1(x)$
and $\bar b_1(x) \ne b_1(x)$, as predicted for example by meson cloud models~\cite{Paiva:1996dd}.

%============================================
\subsection{Intrinsic heavy quark PDFs from non-singlet evolution}
\label{sec:PDFdef}
%============================================
The initial $x$-dependence at the scale of the charm mass is defined via Eq.~\eqref{eq:bhps} and the normalization is fixed to match the one predicted by the CTEQ6.6c0 fit. The IB PDF was generated using the {\it Same Scales} boundary conditions (see above) together with the same $x$-dependent input of Eq.~\eqref{eq:bhps}. Both PDFs were then evolved according to the non-singlet evolution equation and the corresponding grids were produced. Note that because in our approximation, the evolution of the intrinsic charm and bottom PDFs is completely decoupled, the normalization of our PDFs can be easily changed by means of simple rescaling.

%============================================
\subsection{Numerical validation}
\label{sec:tests}
%============================================
We use the CTEQ6.6c series of intrinsic charm fits to test the ideas presented above, and in particular CTEQ6.6c0 and CTEQ6.6c1 which employ the BHPS model with $1\%$ and $3.5\%$ IC probability, respectively.\footnote{This corresponds to the values of 0.01 and 0.035 of the first moment of the charm PDF, $\int dx  \, c(x)$, calculated at the input scale $Q_0=m_c=1.3\ \GeV$.} In the following we compare CTEQ6.6c0 and CTEQ6.6c1 sets where IC has been obtained from global analysis to our approximate IC PDFs supplemented with the central CTEQ6.6 fit, which has a radiatively generated charm distribution.
\begin{figure}
\begin{center}
\subfigure[]{
\label{fig-testa}
\includegraphics[angle=0,width=0.46\textwidth]{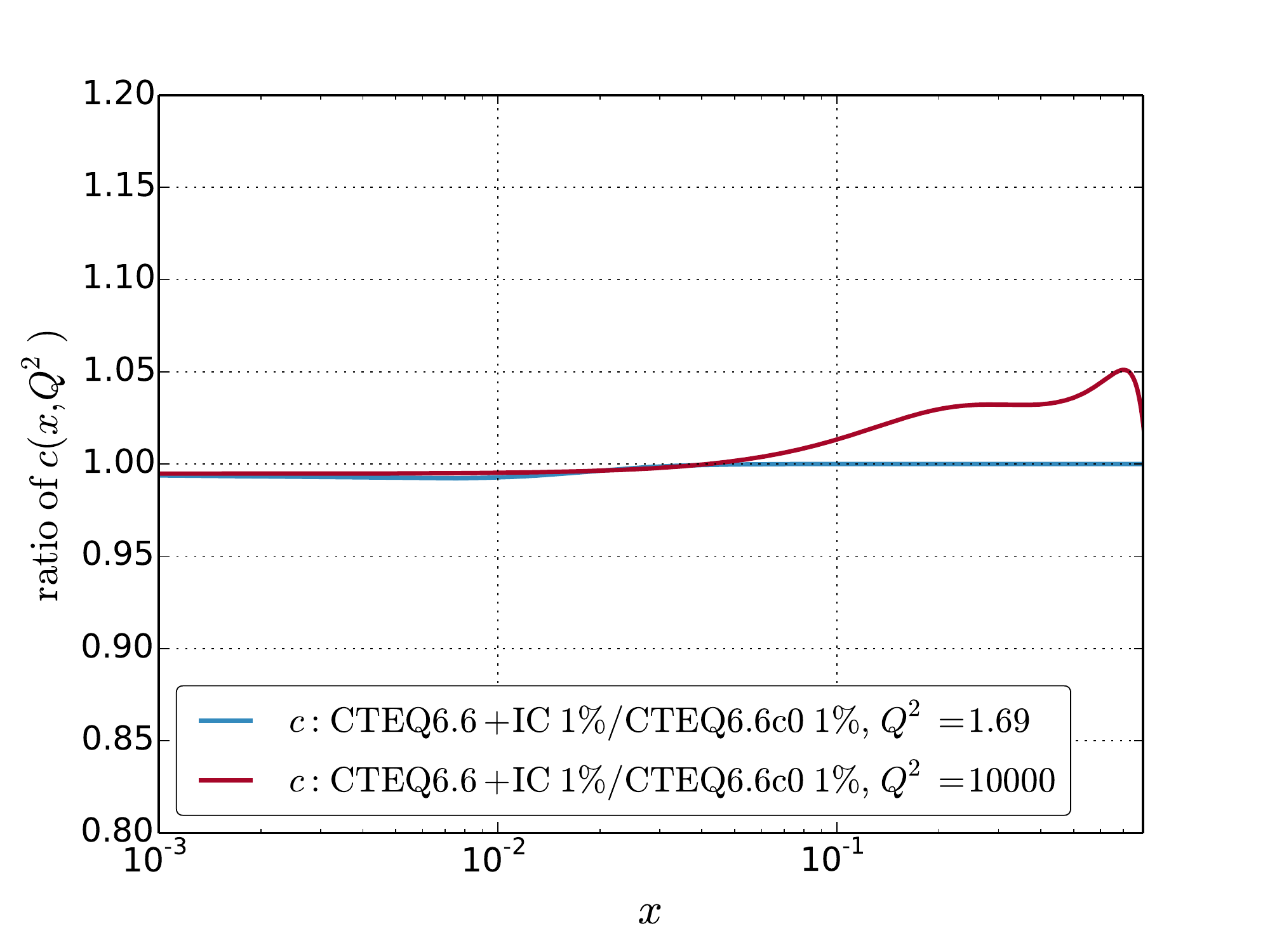}}
\subfigure[]{
\label{fig-testb}
\includegraphics[angle=0,width=0.46\textwidth]{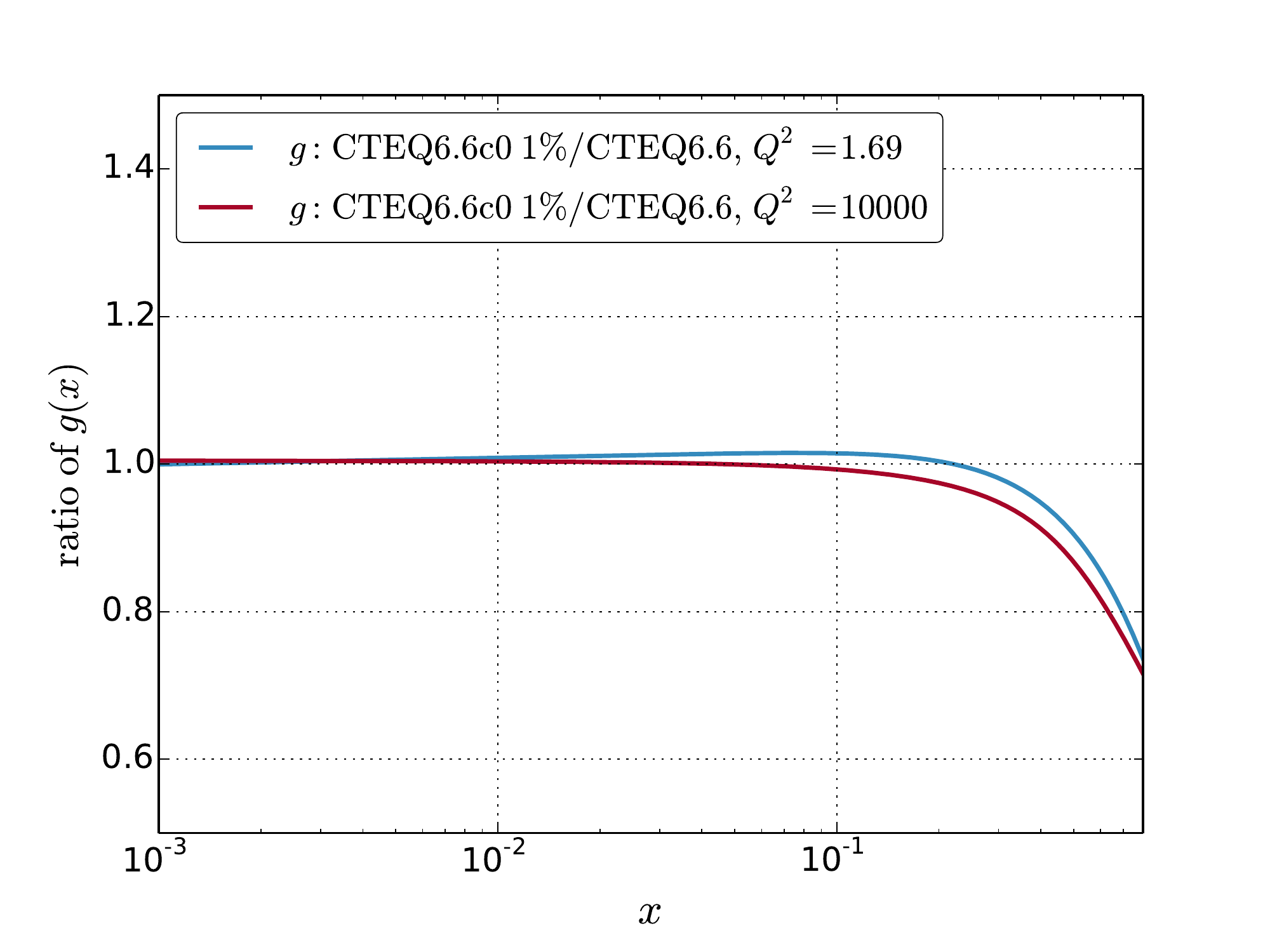}}
\caption{(a) Ratio of the CTEQ6.6 + IC 1\% and CTEQ6.6c0 (1\%) charm distributions. (b) Ratio of the CTEQ6.6c0 and the CTEQ6.6 gluon distributions. The results are shown as function of $x$ for two scales, $Q^2=1.69$ and $Q^2=10000\ \GeV^2$.}
\end{center}
\end{figure}

In Fig.~\ref{fig-testa}, one can see that  at low $Q^2$ the difference between the sum $c_0+c_1$ and the CTEQ6.6c0 charm distribution is tiny, and smaller than $5\%$ at the higher $Q^2$. 

The inclusion of the intrinsic charm distribution will alter the other parton distributions, most notably the gluon PDF. In order to gauge this effect, we compare in Fig.~\ref{fig-testb} the gluon distribution from the standard CTEQ6.6 fit with the one from the CTEQ6.6c0 analysis. For small $x$ ($x<0.1$) the gluon PDF is not affected by the presence of a BHPS-like intrinsic charm component which is concentrated at large $x$. We note that at large-$x$, where most of the difference lies, the gluon distribution is already quite small and the uncertainty of the gluon PDF is sizable (of order of 40 -- 50\% for the CTEQ6.6 set).

We conclude that for most applications,  adding a standalone intrinsic charm distribution to an existing standard global analysis of PDFs is internally consistent and leads only to a small error. Moreover, for the case of intrinsic bottom which is additionally suppressed, the accuracy of the approximation will be even better. For completeness we also provided similar validation using the parton--parton luminosities in~\cite{Lyonnet:2015dca} to which we refer the reader for further details.

%%%%%%%%%%%%%%%%%%%%%%%%%%%%%%%%%%%%%%%%%%%%
\section{LHC observables: possible effects of IC/IB}
\label{sec:numerics}

We study the effects of both IC and IB on parton--parton luminosities,
$\frac{d\mathcal{L}_{ij}}{d\tau}$, at LHC with $\sqrt{S}=14$ TeV.
We define the luminosity as
\begin{equation}
\frac{d\mathcal{L}_{ij}}{d\tau}(\tau,\mu) = \frac{1}{1+\delta_{ij}}\frac{1}{\sqrt{S}}\int_\tau^1 \frac{dx}{x}
                       \Big[ f_i(x,\mu)f_j(\tau/x,\mu) + f_j(x,\mu)f_i(\tau/x,\mu) \Big],
\label{eq:pplumi}
\end{equation}
where $\tau=x_1 x_2$ and $f_i$ are parton distribution functions.
This allows us to assess the relevance of a non-perturbative heavy quark component for the production of new heavy particles coupling to the SM fermions.

\begin{figure}[t]
	\centering
	\includegraphics[width=0.43\textwidth]{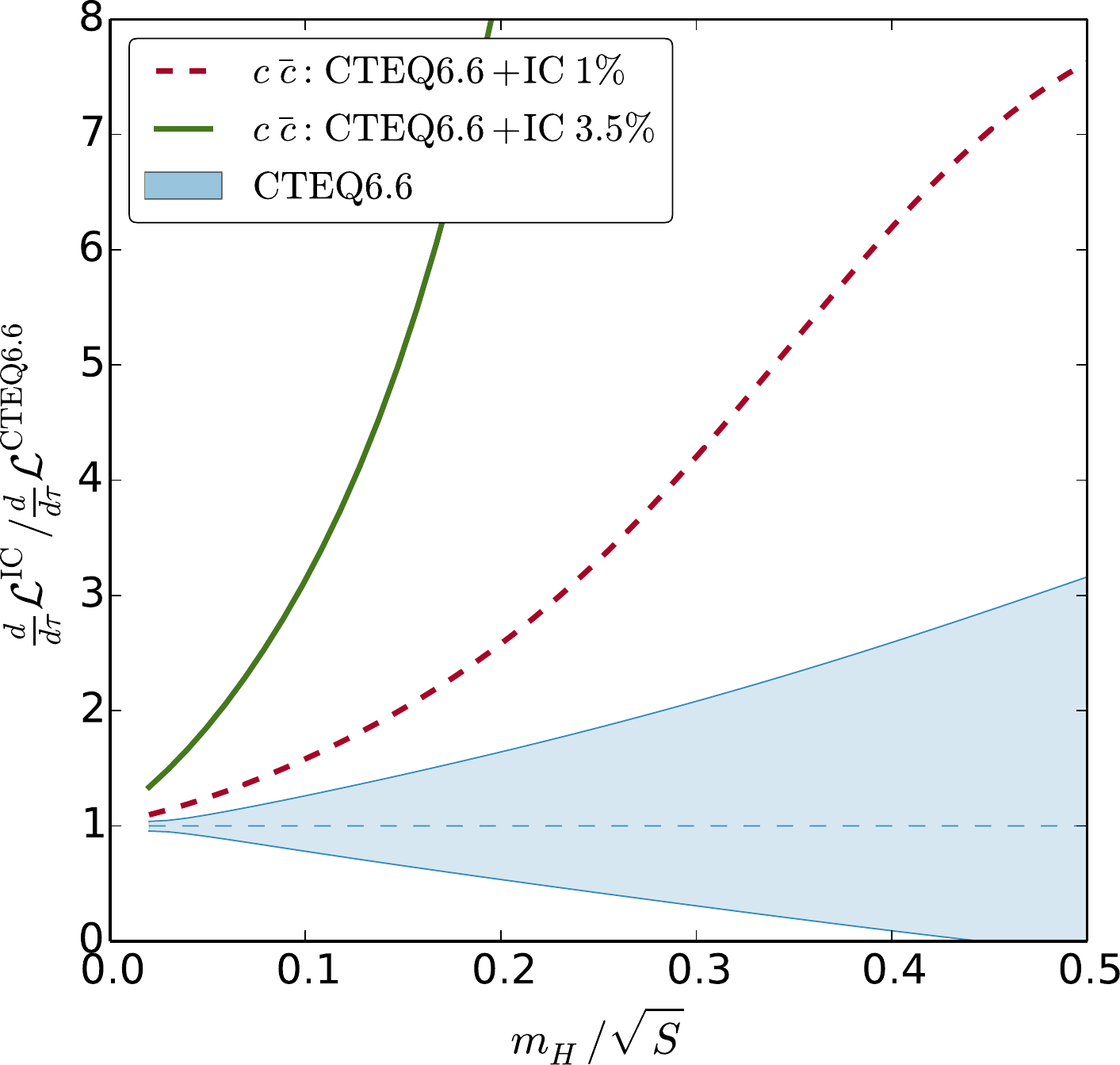}
	\includegraphics[width=0.43\textwidth]{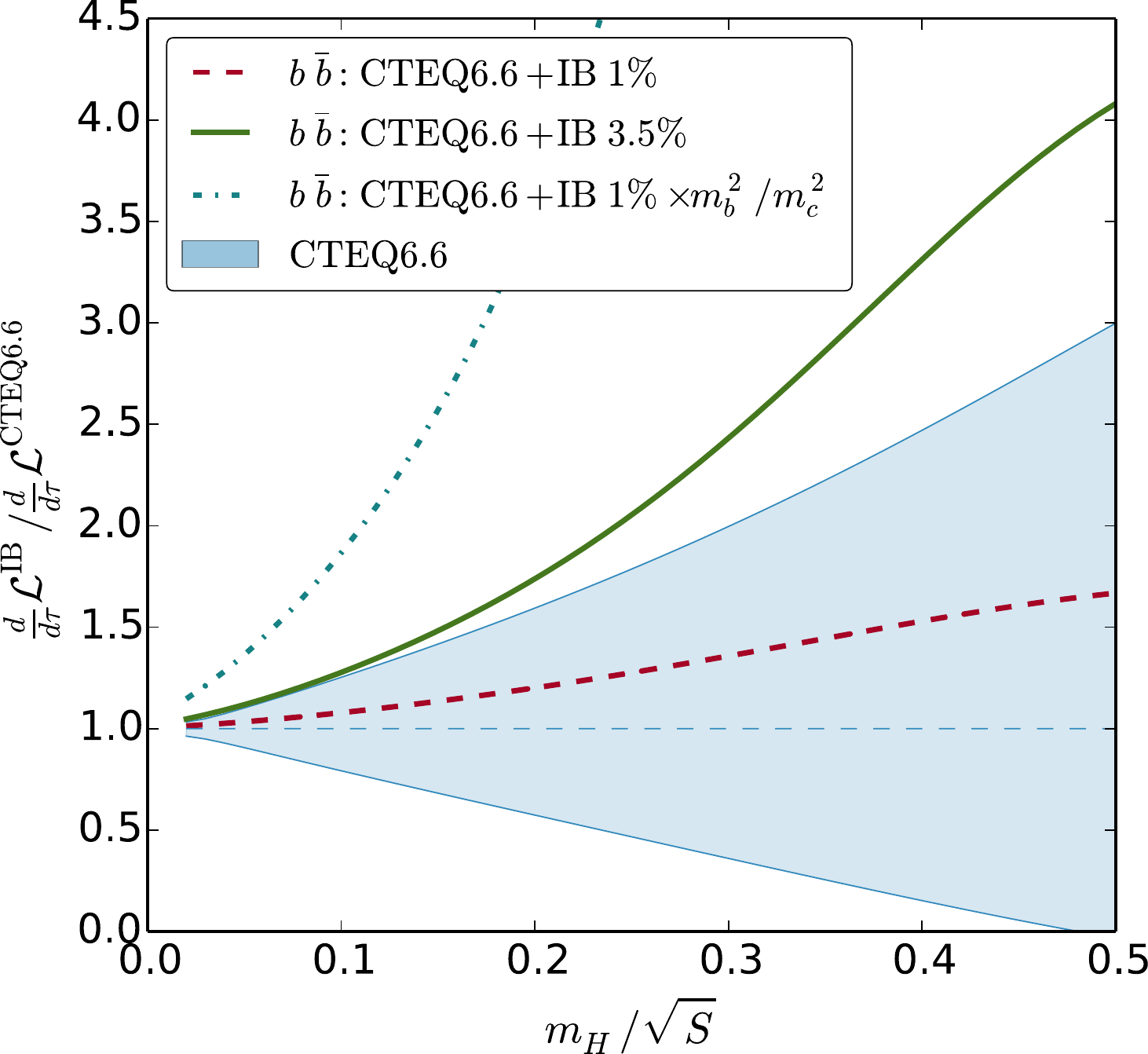}
	\caption{Ratio of $c \bar c$ luminosities (left) and $b \bar b$ luminosities (right) at the LHC14 
	for charm(bottom)-quark PDF sets with and without an intrinsic component as a function of $\sqrt{\tau}=m_H/\sqrt{S}$. 
	The ratio for the $c \bar c$ ($b \bar b)$ luminosity (solid, green line) in the left (right) figure reaches values of 50 (17) at $\sqrt{\tau}=0.5$. In addition to the curves with 1\% normalization (red, dashed lines) we include the results for the 3.5\% normalization (green, solid lines)
which was found to be still compatible with the current data~\cite{Nadolsky:2008zw}.
	}
	\label{fig-ratio_w_uncertainty_c}
\end{figure}

To explore how the presence of IC and IB would affect physics observables with a non-negligible heavy quark initiated subprocesses, in Fig.~\ref{fig-ratio_w_uncertainty_c} we show the ratios of luminosities for charm and bottom with and without an intrinsic contribution for 1$\%$ and 3.5$\%$ normalizations. Furthermore, since there are no experimental constraints on the IB normalization, in Fig.~\ref{fig-ratio_w_uncertainty_c} (right), we also include an extreme scenario where we remove the usual  $m_c^2/m_b^2$ factor; thus, the first moment of the IB is 1$\%$ at the initial scale $m_c$.
As expected, in the case of IB the effect is smaller. However, for the $b \bar{b}$ luminosity a 3.5$\%$ normalization of IB leads to a curve which lies clearly above 
the error band of the purely perturbative result. 
In the extreme scenario (which is not likely but by no means excluded) the IB component has a big effect on the $b \bar{b}$. 

Note that similar results for the $c g$ and $b g$ luminosities have been obtained but are not shown here, see~\cite{Lyonnet:2015dca}.

%%%%%%%%%%%%%%%%%%%%%%%%%%%%%%%%%%%%%%%%%%%%
\section{Conclusions}
\label{sec:conclusions}
%%%%%%%%%%%%%%%%%%%%%%%%%%%%%%%%%%%%%%%%%%%%

This contribution presented a new method to generate a matched IC/IB distributions for any PDF set without the need for a complete global re-analysis. This renders easy to carry out a consistent analysis including intrinsic heavy quark effects. In addition, because the evolution equation for the intrinsic heavy quarks decouples, one can freely adjust the normalization of the IC/IB PDFs.

We showed that our approximation holds to a very good precision for the IB. For the IC, the error is larger (because the IC increases), yet our method is still useful. Indeed, for an IC normalization of  1-2\%, the error is less than the  PDF uncertainties at the large-$x$ where the IC is relevant. If the normalization is larger, although the error may be the same order as the PDF uncertainties, the IC effects also grow and can be separately distinguished from the case without IC. In any case, the IC/IB represents a non-perturbative systematic effect which should be accounted for.

The method presented here greatly simplifies our ability to estimate the impact of the intrinsic heavy quark effects on the new physics searches. It can also be very useful in searching for and constraining the intrinsic charm and bottom components of the nucleon by itself. In particular in the future facilities such as an Electron Ion Collider (EIC), the Large Hadron-Electron collider (LHeC), or AFTER\at LHC.

The PDF sets for intrinsic charm and intrinsic bottom discussed in this analysis
(1\% IC, 3.5\% IC, 1\% IB, 3.5\% IB) are available from the authors upon request.

\bibliographystyle{woc}
\bibliography{iQ} 
% BibTeX or Biber users please use (the style is already called in the class, ensure that the "woc.bst" style is in your local directory)
% \bibliography{name or your bibliography database}

\end{document}